\documentclass[preprint,aps]{revtex4}

\begin{document}

\title{The Extended Equivalence Principle and the Kramers-Kronig Relations}
\author{Raymond Y. Chiao}
\date{September 30, 2002 version of PRL (REVTEX) MS}

\begin{abstract}
A seemingly obvious extension of the weak equivalence principle, in which
all matter must respond to Post-Newtonian gravitational fields, such as
Lense-Thirring and radiation fields, in a \textit{composition-independent}
way, is considered in light of the Kramers-Kronig dispersion relations for
the linear response of any material medium to these fields. \ It is argued
that known observational facts lead to violations of this extended form of
the equivalence principle. (PACS numbers: 04.80.Cc, 04.80.Nn, 03.65.Ud,
67.40.Bz)
\end{abstract}

\address{Department of Physics\\
University of California\\
Berkeley, CA 94720-7300}

\maketitle

The equivalence principle lies at the foundation of General Relativity (GR);
similarly, the superposition principle lies at the foundation of Quantum
Mechanics (QM). \ There exist a profound conceptual tension between these
two principles, which originates from the clash between the notion of $%
locality$ contained within the equivalence principle, and the notion of $%
nonlocality$ contained within the superposition principle \cite{ChiaoWheeler}%
. \ In GR, all physical systems are to be viewed as being completely \textit{%
spatially separable} into independent parts, whereas in QM, there exist
entangled states (i.e., sums of product states), which lead to an intrinsic 
\textit{spatial nonseparability} of certain physical systems, as evidenced
by the observed violations of Bell's inequality for these systems \cite%
{Gisin2002}. \ This tension between GR and QM may lead to important
experimental consequences.

Here I shall examine this tension in the restricted context of $weak$
gravity interacting with $slow$ matter, i.e., in the context of the $%
linearized$ Einstein's equations for Post-Newtonian gravitational fields
interacting with $nonrelativistic$ material media, in particular, with
macroscopically coherent quantum-mechanical matter. \ The resulting
Maxwell-like equations lead to Lense-Thirring and gravitational radiation
fields in the near zone (or induction zone) of the source, and allow us to
treat the propagation of gravitational radiation through a medium\ placed in
the near zone of the source \cite{FarZone}. \ 

I shall consider a seemingly obvious extension of the weak equivalence
principle to such Post-Newtonian situations, which shall be called the
``extended equivalence principle,'' i.e., that the \textit{linear response}
of the medium to weak, Post-Newtonian gravitational fields must be \textit{%
independent of the composition or thermodynamic state} of the medium. \ This
extended principle will be considered\ in light of the Kramers-Kronig
relations. \ It leads to a contradiction with some known observational
facts. \ Hence the extended equivalence principle must be a false extension
of the weak equivalence principle.

For weak gravity and slow matter, Maxwell-like equations have been derived
from\ the linearized form of Einstein's field equations \cite{Forward}\cite%
{Braginsky}\cite{Becker}\cite{Tajmar}. \ The electric-like field $\mathbf{E}%
_{G}$, which is to be identified with the local acceleration due to gravity $%
\mathbf{g}$,\ is analogous to the electric field $\mathbf{E}$, and the
magnetic-like induction field $\mathbf{B}_{G}$, which is to be identified
with the Lense-Thirring field, is analogous to the magnetic field $\mathbf{B}
$ of Maxwell's theory. \ The physical meaning of the field $\mathbf{E}%
_{G}\equiv \mathbf{g}$ is that it is the ordinary, local three-acceleration $%
\mathbf{g}$ of a given particle in a system of locally freely-falling test
particles, whose motion is induced by the gravitational field as seen by an
observer in a local inertial frame located at the center of mass of a
measuring system, which coincides with that of the material medium placed in
the near zone of the source.\ \ The physical meaning of the magnetic-like
induction field $\mathbf{B}_{G}$ is that it is the local angular velocity of
an inertial frame centered on this test particle relative to that of the
observer. The Maxwell-like equations in SI units for these weak
gravitational fields in the near zone of the source, upon setting the PPN
(``Parametrized Post-Newton'') parameters \cite{Braginsky} to be those of
general relativity, are%
\begin{equation}
\mathbf{\nabla \cdot D}_{G}=-\rho _{free}  \label{Gauss-like}
\end{equation}%
\begin{equation}
\mathbf{\nabla \times \mathbf{E}_{G}=-}\frac{\partial \mathbf{B}_{G}}{%
\partial t}  \label{Faraday-like}
\end{equation}%
\begin{equation}
\mathbf{\nabla \cdot B}_{G}=0
\end{equation}%
\begin{equation}
\mathbf{\nabla \times H}_{G}\mathbf{=}-\mathbf{j}_{free}+\frac{\partial 
\mathbf{D}_{G}}{\partial t}  \label{Ampere-like}
\end{equation}%
where $\mathbf{D}_{G}$ is a displacement-like field, $\mathbf{H}_{G}$ is a
magnetic-like field intensity, $\rho _{free}$ is the free mass density and $%
\mathbf{j}_{free}$ is the free mass current density\ of \textit{freely
falling} matter, as seen by the observer \cite{Free-Bound-Sources}. \ In the
vacuum just outside the source, the fields $\mathbf{D}_{G}$ and $\mathbf{H}%
_{G}$ are related to fields $\mathbf{E}_{G}$ and $\mathbf{B}_{G}$ by the $%
vacuum$ constitutive relations \cite{BCT}%
\begin{equation}
\mathbf{D}_{G}=4\varepsilon _{G}\mathbf{\mathbf{E}_{G}}
\end{equation}%
\begin{equation}
\mathbf{B}_{G}=\mu _{G}\mathbf{H}_{G}
\end{equation}%
where $\varepsilon _{G}=1/16\pi G=2.98\times 10^{8}$ kg$^{2}/$N$\cdot $m$^{2}
$ and $\mu _{G}=16\pi G/c^{2}=3.73\times 10^{-26}$ m$/$kg \cite{Forward}\cite%
{Braginsky}\cite{Becker}\cite{Tajmar}\cite{Landau}. $\ $\ Here $G$ is
Newton's constant.

For obtaining solutions in the near zone of the source, we convert these
Maxwell-like equations for weak gravity into a wave equation in the standard
way, and conclude that the speed of GR waves in the vacuum near the source
is $c_{G}=\frac{1}{2}(\varepsilon _{G}\mu _{G})^{-1/2}=1.499\times 10^{8}$ m$%
/$s, which is half the vacuum speed of light \cite{Half-c}, and that the
impedance of free space for GR waves in the vacuum near the source is $%
Z_{G}=(\mu _{G}/\varepsilon _{G})^{1/2}=1.12\times 10^{-17}$ m$^{2}/$s$\cdot 
$kg. \ These two quantities would correspond respectively to the vacuum
speed of light $c=(\varepsilon _{0}\mu _{0})^{-1/2}=2.998\times 10^{8}$ m$/$%
s,\ and to the impedance of free space $Z_{0}=(\mu _{0}/\varepsilon
_{0})^{1/2}=377$ ohms in Maxwell's theory. \ Since the forms of these
Maxwell-like equations are identical to those of Maxwell's apart from a sign
change of the sources $-\rho _{free}$ and $-\mathbf{j}_{free}$ in the
Gauss-like and the Ampere-like laws, Eqs.(\ref{Gauss-like}) and (\ref%
{Ampere-like}), respectively, the same boundary conditions follow from them,
and therefore the same unique solutions for corresponding electromagnetic
problems carry over in a unique manner to the gravitational ones in the near
zone.

Let us now place an isotropic medium in the vicinity (i.e., in the near
zone) of the source, and introduce the constitutive relations for this medium%
\begin{equation}
\mathbf{D}_{G}=4K_{E}\varepsilon _{G}\mathbf{\mathbf{E}_{G}}
\label{kappa_GE}
\end{equation}%
\begin{equation}
\mathbf{B}_{G}=K_{M}\mu _{G}\mathbf{H}_{G}  \label{kappa_GM}
\end{equation}%
\begin{equation}
\mathbf{j}_{G}=-\sigma _{G}\mathbf{\mathbf{E}_{G}}  \label{sigma_G}
\end{equation}%
where $K_{E}$ is the gravitational dielectric-like constant of the medium, $%
K_{M}$ is its gravitational magnetic-like relative permeability, and $\sigma
_{G}$ is the gravitational analog of the electrical conductivity of the
medium, whose magnitude is inversely proportional to its viscosity. \ Due to
the sign change in the sources $-\rho _{free}$\ and $-\mathbf{j}_{free}$ of
the Gauss-like and Ampere-like laws, it is natural to choose to define the
third constitutive relation with a minus sign, so that for $dissipative$
media, $\sigma _{G}$ is always a $positive$ quantity. \ That this third
constitutive relation should be introduced here is motivated by the fact
that gravitational radiation produces a quadrupolar \textit{shear field}, so
that one might expect that the \textit{shear viscosity} of the medium
through which it propagates, should enter into the dissipation of the wave.
\ Otherwise, there could never be any dissipation of a gravitational wave as
it propagates through any medium. \ The phenomenological parameters $K_{E},$ 
$K_{M},$ and $\sigma _{G}$ must be determined by experiment, just like the
corresponding electromagnetic quantities, viz., the dielectric constant $%
\kappa _{e}$, the relative magnetic permeability $\kappa _{m}$, and the
electrical conductivity $\sigma _{e}$, in Maxwell's theory. \ 

But why is it even $permissible$ to introduce nontrivial constitutive
relations, Eqs.(\ref{kappa_GE}), (\ref{kappa_GM}), and (\ref{sigma_G}), with 
$K_{E}\neq 1$, $K_{M}\neq 1$, and $\sigma _{G}\neq 0$? \ One argument $%
against$ the introduction of such relations is that the sources of all
gravitational fields are entirely determined by the masses and the mass
currents of the material medium in a \textit{composition-independent} way,
since the source of spacetime curvature in Einstein's field equations arises
solely from the stress-energy tensor, which includes the energy density,
momentum density, and stress associated with $all$ forms of matter and $all$
nongravitational fields \cite{MTW}. \ Therefore, the source of spacetime
curvature must be independent of the specific composition of the source. \
Likewise, the response of all matter to spacetime curvature must also be
independent of the specific composition of the matter. \ By ``composition
independence,'' we mean here not only the independence from the specific
kinds of chemical, nuclear, and elementary particle content of the matter in
question (including all of the specific kinds of nongravitational
interactions which bind that matter together), but also the independence
from the specific thermodynamic state of the matter. \ For example, a drop
of liquid water, or a particle of frozen ice, would both undergo exactly the
same geodesic, free-fall motion in the Earth's gravitational field for the
same initial conditions. \ In this line of reasoning, there cannot be any
mysterious property of the material which would make $K_{E}$, $K_{M}$, and $%
\sigma _{G}$\ different for different materials. \ The only permissible
values of these constants would then be their vacuum values, viz., $K_{E}=1$%
, $K_{M}=1$, and $\sigma _{G}=0$. \ Otherwise, there would exist a seeming
violation of the equivalence principle, since there would then exist a 
\textit{composition-dependent} response of different kinds of materials to
gravitational fields.

However, one must carefully distinguish between the $weak$ equivalence
principle, where the composition independence of the \textit{free-fall
response} of matter to gravity has been extensively experimentally tested in
the low-frequency, Newtonian-gravity limit, and the $extended$ equivalence
principle, which extends the composition independence of the weak
equivalence principle to include\ the \textit{linear response} of all kinds
of matter to Post-Newtonian gravitational fields, such as Lense-Thirring and
gravitational radiation fields. \ Because of the difficulty of generating
and detecting these Post-Newtonian fields, this extended form of the
equivalence principle has not been extensively experimentally tested.

The above Maxwell-like equations, in conjunction with the constitutive
relations, Eqs.(\ref{kappa_GE}), (\ref{kappa_GM}), and (\ref{sigma_G}), with 
$K_{E}\neq 1$, $K_{M}\neq 1$, and $\sigma _{G}\neq 0$, lead to the near-zone
propagation in the medium of a gravitational plane wave at a frequency $%
\omega $ at a phase velocity $v_{phase}(\omega )$ given by%
$v_{phase}(\omega )=c_{G}/n_{G}(\omega ),$
where the index of refraction $n_{G}(\omega )$ of the medium is given by%
$n_{G}(\omega )=\left( K_{E}(\omega )K_{M}(\omega )\right) ^{1/2}.$
Such a solution is formally identical to that for an electromagnetic plane
wave propagating inside a dispersive optical medium, whose index of
refraction is given by $n(\omega )=\left( \kappa _{e}(\omega )\kappa
_{m}(\omega )\right) ^{1/2}$. \ Due to the validity of the linearity
approximation in the weak gravity limit, induction-zone fields can always be
Fourier decomposed into a superposition of such plane waves, which form a
complete basis set. \ 

Since we are considering the\ \textit{linear response} of the medium to $%
weak $ gravitational radiation fields, and since the response of the medium
must be $causal$, the index of refraction $n_{G}(\omega )$ must obey the
Kramers-Kronig relations \cite{Ditchburn}%
\begin{equation}
\mbox{\rm{Re} }n_{G}(\omega )-1=\frac{1}{\pi }P\int_{-\infty }^{+\infty }%
\frac{\mbox{\rm{Im} }n_{G}(\omega ^{\prime })}{\omega ^{\prime }-\omega }%
d\omega ^{\prime }
\end{equation}%
\begin{equation}
\mbox{\rm{Im} }n_{G}(\omega )=-\frac{1}{\pi }P\int_{-\infty }^{+\infty }%
\frac{\mbox{\rm{Re} }n_{G}(\omega ^{\prime })-1}{\omega ^{\prime }-\omega }%
d\omega ^{\prime },  \label{K-K2}
\end{equation}%
where $P$ denotes the Cauchy Principal Value. \ The zero-frequency sum rule
follows from the first of these relations, viz.,%
\begin{equation}
\mbox{\rm{Re} }n_{G}(\omega \rightarrow 0)=1+\frac{c}{\pi }\int_{0}^{+\infty
}\frac{\alpha _{G}(\omega ^{\prime })}{\left( \omega ^{\prime }\right) ^{2}}%
d\omega ^{\prime },  \label{Zero-Frequency-Sum-Rule}
\end{equation}%
where $\alpha _{G}(\omega )$ is the power attenuation coefficient of the
gravitational plane wave at frequency $\omega $ propagating through the
medium, i.e., $\exp (-\alpha _{G}(\omega )z)$ is the exponential factor
which attenuates the power of a wave propagating along the $z$ axis. \ For
media in the ground state, the integrand on the right-hand side of this
zero-frequency sum rule is always positive definite. \ A nonvanishing
dissipation coefficient $\sigma _{G}(\omega )$ introduced in conjunction
with the third constitutive relation will lead to a nonvanishing, positive
value of $\alpha _{G}(\omega )$. \ Since the medium is in its ground state,
the gravity wave cannot grow exponentially with propagation distance $z$. \
Hence $\alpha _{G}(\omega )>0$ for all frequencies $\omega $, and therefore $\mbox{\rm{Re} }n_{G}(\omega \rightarrow 0)>1.$

Now suppose that the extended equivalence principle were true. \ Then the
response of any medium to gravitational radiation at all frequencies must be
characterized by $K_{E}(\omega )=1$ and $K_{M}(\omega )=1$, independent of
the composition of the medium. \ In particular at low frequencies, this
implies that the index of refraction $n_{G}(\omega \rightarrow 0)=1  \label{n_G(0)=1}$ 
must strictly\ be unity. \ It then follows from the above zero-frequency sum
rule, that the attenuation coefficient $\alpha _{G}(\omega )=0  \label{alpha=0}$ must strictly vanish for all frequencies $\omega $. \ This result would
imply that $any$ absorption of gravitational radiation in the near zone of
the source would be impossible at $any$ frequency by $any$ kind of matter. \
Detectors of gravitational radiation would be impossible. \ By reciprocity,
emission of gravitational radiation by any kind of matter at any frequency
would likewise also be impossible. \ Gravitational radiation might as well
not exist \cite{Loinger}. \ This, however, is contradicted by the
observations of Taylor and Weisberg \cite{Taylor}. \ Thus the extended
equivalence principle must be a false extension of the weak equivalence
principle.

Since we are dealing with the linear response of a medium placed in the%
\textit{\ near zone} of the source, it may be objected that the concept of 
\textit{index of refraction}, which usually refers to a medium placed in the
far zone (or radiation zone) of the source, has been used in the above
argument. \ In fact, the use of the far-zone approximation is not necessary
here; the concept of index of refraction $can$ be a valid one for the
treatment of the interaction of a medium with radiation fields in the near
zone of a planar source, for example, in the response of a thin, dissipative
film to a plane wave, when the film is placed parallel to this planar source
and is separated from it by a distance much less than a wavelength \cite%
{ChiaoWheeler}. \ 

Moreover, it is possible to recast the above argument in terms of the
Kramers-Kronig relations for the real and imaginary parts of the
dielectric-like constant $K_{E}$ and of the relative permability-like
constant $K_{M}$, which are valid concepts for the treatment of the
near-zone response of the medium. \ No concept of \textit{wave propagation}
is necessarily involved here when one only uses $K_{E}$ and $K_{M}$
separately in the treatment of the medium's $linear$ and $causal$ response
to Post-Newtonian gravitational fields. \ There result from the separate
Kramers-Kronig relations for $K_{E}$ and $K_{M}$ the two separate
zero-frequency sum rules \cite{Landau2}\cite{Jackson}%
\begin{equation}
\mbox{\rm{Re} }K_{E}(\omega \rightarrow 0)=1+\frac{2}{\pi }\int_{0}^{+\infty
}\frac{\mbox{\rm{Im} }K_{E}(\omega ^{\prime })}{\omega ^{\prime }}d\omega
^{\prime },
\end{equation}%
\begin{equation}
\mbox{\rm{Re} }K_{M}(\omega \rightarrow 0)=1+\frac{2}{\pi }\int_{0}^{+\infty
}\frac{\mbox{\rm{Im} }K_{M}(\omega ^{\prime })}{\omega ^{\prime }}d\omega
^{\prime },
\end{equation}%
where $\omega \mbox{\rm{Im} }K_{E}(\omega )$ is proportional to the power
absorption coefficient due to the imaginary part of $K_{E}(\omega )$, and $%
\omega \mbox{\rm{Im} }K_{M}(\omega )$ is proportional to the power
absorption coefficient due to the imaginary part of $K_{M}(\omega )$. \
Again, for a medium in the ground state, the integrands on the right-hand
sides are positive definite. \ The above argument repeated for these sum
rules leads to the absurd conclusion that there is $no$ possibility for $any$
absorption of energy by $any$ medium in its response to gravitational fields
at $any$ frequency, independent of the composition of the medium. \ 

\ Hence, not only is it $permissible$, but it is also $necessary$, to
introduce constitutive equations with nontrivial, composition-dependent
values of $K_{E}$, $K_{M}$, and $\sigma _{G}$. \ In the case of
electromagnetism, we know that nontrivial values of the relative
permeability $\kappa _{m}\neq 1$ are only possible due to purely
quantum-mechanical effects, viz., paramagnetism and ferromagnetism, due to
the alignment of electron spin, and diamagnetism, such as the Meissner
effect in superconductors, due to London's rigidity of the macroscopic
condensate wavefunction of Cooper pairs of electrons. \ Likewise, one
expects here that nontrivial values of the gravitational relative
permeability $K_{M}\neq 1$ can also only be possible due to purely
quantum-mechanical effects.

There may already be experimental evidence of the existence of a strong
Meissner-like effect in the case of superfluid helium, which can be
interpreted as evidence of a gross, quantum violation of the extended
equivalence principle. \ The Hess-Fairbank effect \cite{Hess-Fairbank} is an
analog of the Meissner effect observed in a slowly rotating bucket of liquid
helium, as it is slowly cooled through the normal-superfluid (or lambda)
transition temperature, in which the angular momentum of the rotating normal
fluid is $expelled$ from the superfluid. \ This may be viewed as evidence
for a true Meissner-like effect, in which the Lense-Thirring field arising
from distant matter in the Universe is expelled. \ From an application of a
limited form of Mach's principle in GR using the Lense-Thirring effect from
distant, rotating cylindrical shells of stars \cite{MTW}, it follows that
the parabolic shape of the surface of any steadily rotating bucket of $any$
classical fluid, after it has reached mechanical equilibrium, is described
by $y=\frac{1}{2}\Omega ^{2}x^{2}/g,$ where $x$ is the distance from the bucket's axis of rotation of a fluid
element on the surface, $y$ is the height of this fluid element with respect
to the height of the liquid surface at $x=0$, $\Omega $ is the bucket's rate
of rotation with respect to the fixed stars, and $g$ is the acceleration due
to Earth's gravity. \ This result, which is a consequence of the equivalence
principle, is independent of the composition of the liquid. \ For example,
steadily rotating buckets filled with mercury, or water, or normal liquid
helium above its superfluid transition temperature, will\ all come to the
same parabolic shape of surface for the same $\Omega $, independent of
atomic composition. \ 

One implication of the Hess-Fairbank effect is that the parabolic surface of
a bucket of rotating normal liquid helium should become flat below the
thermodynamic transition from the normal to the superfluid state. \ It is
possible to measure the curvature of the surface of the rotating fluid by
reflecting a laser beam from it. \ The flattening of the surface below the
lambda transition temperature of liquid helium in a slowly rotating bucket,
as it is cooled slowly enough so that it always stays in thermal equilibrium
throughout the normal-to-superfluid thermodynamic phase transition, has been
observed by\ the use of laser optical systems and laser interferometry \cite%
{FairbanksLaser}. \ Since the flattened-surface response of $superfluid$
helium to the Lense-Thirring field of the distant matter of the Universe is
clearly different from the\ parabolic-surface response of any $classical$
fluid, this effect should be viewed as evidence for a macroscopic quantum
violation of the extended equivalence principle.

For all of the above reasons, one concludes that the extended equivalence
principle is an invalid generalization of the weak equivalence principle.

I would like to thank J. C. Garrison, J. M. Leinaas, C. F. McCormick, R. L.
Packard, and A. D. Speliotopoulos for helpful discussions. \ This paper has
been extracted from a Chapter of a Volume dedicated to John Archibald
Wheeler, which was supported by the John Templeton Foundation. \ This work
was also supported by the late Y. F. Chiao, and by the ONR.


\begin{references}
\bibitem{FarZone} However, this is not to exclude the possibility of a generalization of these results to a medium placed in the far field (i.e., the radiation zone) of the source.
\bibitem{Free-Bound-Sources} In analogy with Maxwell's theory, where one must
carefully distinguish between $free$ and $bound$ charges and charge currents as sources,
one must carefully distinguish here also between $free$ and $bound$ masses and
mass currents as sources, where ``free'' here means ``freely falling.''  It is
to be emphasized that in the Gauss-like and Ampere-like equations, Eqs.(\ref%
{Gauss-like}) and (\ref{Ampere-like}), the sources here refer to $free$ masses
and $free$ mass currents.
\bibitem{BCT}These vacuum constitutive relations are due to the corrections 
to the Newtonian approximation when the motion of the generating mass is taken 
into account (see [5]).
\bibitem{Half-c} One gets $c/2$ for the wave speed because the PPN limit 
assumes that any velocity in the system is much smaller than $c$ (see [11]).  
  The PPN approximation can thus incorporate the generation of gravitational waves (GW)
by any slowly moving mass in the induction zone.  However, since it assumes that any 
velocities within the mass 
moves slowly compared to $c$, and since GWs propogate at $c$, 
it cannot consistently incorporate GWs into its formalism in the radiation zone, but 
only in the induction zone, where the metric is overwhelmingly determined by the 
generating mass.  
However, the correct value of $c$ for radiation-zone 
GW propagation comes out of a similar
set of vacuum Maxwell-like equations, which have been derived 
without making the PPN approximation 
(I thank A. D. Speliotopoulos for pointing this out to me).
\end{references}
\end{document}